\documentclass{jpsj2}

\title{Heat conduction in a 1D harmonic chain with three dimensional vibrations}

\author{Zonghua  \textsc{Liu}$^{1,2}$\thanks{E-mail address: zhliu@phy.ecnu.edu.cn},
Baowen \textsc{LI}$^{2,3}$\thanks{E-mail address: phylibw@nus.edu.sg} }

\inst{$^{1}$Institute of Theoretical Physics and Department of
Physics, East China Normal University, Shanghai, 200062, China\\
$^{2}$Department of Physics, Centre for Computational Science
and Engineering, National University of Singapore, 117542 Singapore\\
$^{3}$NUS Graduate School for Integrative Sciences and
Engineering, Singapore 117597, Republic of Singapore}

\recdate{\today}
\abst{We study vibrational energy transport in a quasi 1-D harmonic chain
with both longitudinal and transverse vibrations. We demonstrate via
both numerical simulation and theoretic analysis that for 1-D atomic chain connected
by 3D harmonic springs, the coefficient of heat conduction
changes {\it continuously} with its lattice constant, indicating the
qualitative difference from the corresponding 1-D case where the
coefficient is independent of the lattice constant.}

\kword{heat conduction; harmonic chain; 3D vibration; lattice
constant; energy transport}

\begin{document}
\maketitle

\section{Introduction}

Heat conduction in low dimensional systems (less than
three dimension) has been attracting increasing attention in the
past decade\cite{review}. From the fundamental point of view, one
would like to know whether the fundamental transport theory for
bulk material, such as the Fourier law of heat conduction, is
still valid for such low dimensional systems. In fact, the
question is not trivial, as there is still no rigorous proof
available so far. From application point of view, it is an
indispensable question to understand the heat conduction
properties of the nanoscale materials before they are put into
application \cite{nano}.

In order to understand the underlying physical mechanism of heat
conduction in one dimensional (1D) systems, different lattice
models with and without on-site (pinning) potential have been
used, such as the Fermi-Pasta-Ulam model \cite{FPU} without
on-site potential, the Frenkel Kontorova model\cite{FK}, and the
$\phi^4$ model with on-site potential \cite{F4}. With these
models, the roles of anharmonicity and the on-site potential in
heat conduction in 1D systems have been nicely demonstrated. For
example, in the system without on-site potential, a size-dependent
thermal conductivity has been observed \cite{FPU} due to a
superdiffusive motion of phonons \cite{LW03}. In the system with
on-site potential, the scattering from on-site potential makes
phonon transport diffusively, leading to a size-independent
thermal conductivity. Considering the fact that in the 1D models,
the lattice is restricted to longitudinal vibration only, i.e.,
the transverse motions have been completely ignored, we conclude
that these 1D models are too simple to be used for modelling heat
conduction in realistic nanostructures, such as nanotube and
nanowires.

Indeed, the heat conduction behavior changes when the transverse
vibrations are considered. In a recent work, Wang and
Li\cite{WangLi04} considered a model of quasi 1D chain connected
by 2D spring with a bending angle interaction. The lattices are
allowed to vibrate in longitudinal as well as in one transverse
direction. The results show that for a fixed lattice constant, the
anomalous heat conduction coefficient $\kappa(N)$ changes from a
logarithmic divergence with system size $N$, i.e, $log N$ for
large transverse coupling, to a $1/3$ power law divergence,
$N^{1/3}$, at intermediate coupling, and then to a $2/5$ power law
divergence, $N^{2/5}$ at low temperatures and weak coupling. The
results are mainly due to the mode coupling between the
longitudinal modes and transverse modes.

In this paper, we study the quasi 1D chain with motion in both the
longitudinal direction and the two transverse directions. Namely,
the lattices can vibrate in all three directions, which is more
closer to real nano-scale quasi 1D systems. For simplicity, we
call our system a 3D harmonic chain. To be more practical, we will
not going to discuss the thermodynamic limit (length goes to
infinity) as most of the nanoscale systems are of finite length.
Instead, we shall focus on the effect of the lattice constant. The
model discussed by Wang and Li\cite{WangLi04} can be considered as
a simplified polymer chain, while the model studied in our current
paper can be considered as simplified nanotube or nanowire model.

We shall demonstrate soon that, in 1D case, the lattice constant
does not play any role when the atoms can be considered as moving
around its equilibrium; while in 3D case, the situation is totally
different and the lattice constant will influence the heat
conduction. It goes back to the 1D case when the lattice constant
$a\rightarrow \infty$.

\section{Model and numerical results}

The 3D harmonic chain is coupled with the Nose-Hoover thermostats
\cite{Nose1984} on the first and last particle, keeping them at
temperature $T_h$ and $T_l$, respectively. The Hamiltonian is
\begin{equation}\label{eq:FPU}
 H=\frac{1}{2}\sum_i p_{ix}^2+p_{iy}^2+p_{iz}^2+\frac{1}{2}\sum_i(|{\bf r}_{i+1}-{\bf r}_i|-a)^2,
\end{equation}
where $a$ is the lattice constant, ${\bf r}_i=(ia+x_i, y_i, z_i)$,
and $x_i, y_i, z_i$ are the displacements from the equilibrium
position $(ia, 0, 0)$. The motion of the particles satisfy the
canonical equations $\dot{q}_{iw}=\partial H/\partial p_{iw}$ and
$\dot{p}_{iw}=-\partial H/\partial q_{iw}$ with $w=x,y,z$,
respectively, and $i=2,3,\cdots,N-1$. The left and right heat
baths are described respectively by the equations,

\begin{eqnarray}
\dot{\xi}_{h}=\frac{1}{3T_h}(\dot{q}_{1x}^2+\dot{q}_{1y}^2+\dot{q}_{1z}^2)-1,
\nonumber\\
\dot{\xi}_{l}=\frac{1}{3T_l}(\dot{q}_{Nx}^2+\dot{q}_{Ny}^2+\dot{q}_{Nz}^2)-1,
\label{eq:HeatBath}
\end{eqnarray}

The equations of the motion for the first and last particle are
$\dot{q}_{1w}=\partial H/\partial p_{1w}$, $\dot{p}_{1w}=-\partial
H/\partial q_{1w}-\xi_h p_{1w}$, $\dot{q}_{Nw}=\partial H/\partial
p_{Nw}$, and $\dot{p}_{Nw}=-\partial H/\partial q_{Nw}-\xi_l
p_{Nw}$ with $w=x,y,z$, respectively. Substituting Eq.
(\ref{eq:FPU}) into the canonical equations we can get the
expressions of $\dot{p}_{ix}$, $\dot{p}_{iy}$ and $\dot{p}_{iz}$
as follows
\begin{eqnarray}\label{eq:pix1}
\dot{p}_{ix}&=& (1-\frac{a}{d_{i+1}})(a+\Delta
x_{i+1})-(1-\frac{a}{d_i})(a+\Delta
x_i), \nonumber\\
\dot{p}_{iy}&=&(1-\frac{a}{d_{i+1}})\Delta
y_{i+1}-(1-\frac{a}{d_i})\Delta
y_i, \\
\dot{p}_{iz}&=&(1-\frac{a}{d_{i+1}})\Delta
z_{i+1}-(1-\frac{a}{d_i})\Delta z_i, \nonumber
\end{eqnarray}
where
$d_{i+1}=\sqrt{(a+\Delta x_{i+1})^2+(\Delta y_{i+1})^2+(\Delta z_{i+1})^2}$,
$\Delta x_{i+1}=x_{i+1}-x_i$, $\Delta y_{i+1}=y_{i+1}-y_i$, and
$\Delta z_{i+1}=z_{i+1}-z_i$.

Re-scaling $q_{iw}$ and $p_{iw}$ as $q'_{iw}=q_{iw}/a$ and
$p'_{iw}=p_{iw}/a$ with $w=x,y,z$, respectively, Eq.
(\ref{eq:pix1}) becomes
\begin{eqnarray}\label{eq:scall}
\dot{p'}_{ix}&=& (1-\frac{1}{d'_{i+1}})(1+\Delta
x'_{i+1})-(1-\frac{1}{d'_i})(1+\Delta
x'_i), \nonumber\\
\dot{p'}_{iy}&=&(1-\frac{1}{d'_{i+1}})\Delta
y'_{i+1}-(1-\frac{1}{d'_i})\Delta
y'_i, \\
\dot{p'}_{iz}&=&(1-\frac{1}{d'_{i+1}})\Delta
z'_{i+1}-(1-\frac{1}{d'_i})\Delta z'_i, \nonumber
\end{eqnarray}
where
$d'_{i+1}=\sqrt{(1+\Delta x'_{i+1})^2+(\Delta y'_{i+1})^2+(\Delta z'_{i+1})^2}$.
The heat bath Eqs. (2) is re-scaled as:

\begin{eqnarray}
\dot{\xi}_{h}=\frac{1}{3T'_h}(\dot{q'}_{1x}^2+\dot{q'}_{1y}^2+\dot{q'}_{1z}^2)-1,
\nonumber\\
\dot{\xi}_{l}=\frac{1}{3T'_l}(\dot{q'}_{Nx}^2+\dot{q'}_{Ny}^2+\dot{q'}_{Nz}^2)-1.
\label{eq:HeatBath}
\end{eqnarray}
where $T'_{h,l}=T_{h,l}/a^2$. This means that the equations
$\dot{p}_{1w}$ and $\dot{p}_{Nw}$ for the boundary particles are
invariant providing that the temperature of the two heat baths are
re-scaled as mentioned above. This is interesting. It indicates that in our
system, the model with twice smaller lattice constant is identical
with the one with four times smaller temperature.

For the case of fixed temperatures $T_h$ and $T_l$, the re-scaling
of $q_{1w}$ and $q_{Nw}$ will cause different $\dot{\xi}_h$ and
$\dot{\xi}_l$, namely, the effect of lattice constant will show up
through the heat baths and result in different behaviors of heat
conduction. This can be understood like this. For a given $T_h$
and $T_l$,  the average of the amplitude of $\dot{q}_{iw}$ are
fixed for the heat bath particles. Therefore, larger $a$ means a
relatively smaller $q'_{iw}$. Making a Taylor expansion of
$d'_{i+1}$ with respect to $\Delta x_i', \Delta y_i'$, and $\Delta
z_i'$, from Eq. (\ref{eq:scall}) we see that $\dot{p'}_{ix}$ is
proportional to the first order of $\Delta x'_i$ and
$\dot{p'}_{iy}$ and $\dot{p'}_{iz}$ are proportional to the second
order of $\Delta x'_i$. Therefore, smaller $q'_{iw}$ or larger $a$
will make the transverse motion negligible, indicating that we may
observe a harmonic motion for the larger $a$. Generally, with the
decrease of $a$, the transverse motion will become more and more
important and thus influence the heat conduction more. When
$q'_{iw}$ reaches the magnitude of unity or of lattice constant,
the heat conduction will be saturated, thus we may observe a
platform for small enough $a$.

\section{Heat conductivity versus system size and lattice constant}

\begin{figure}[tb]
\begin{center}
\includegraphics{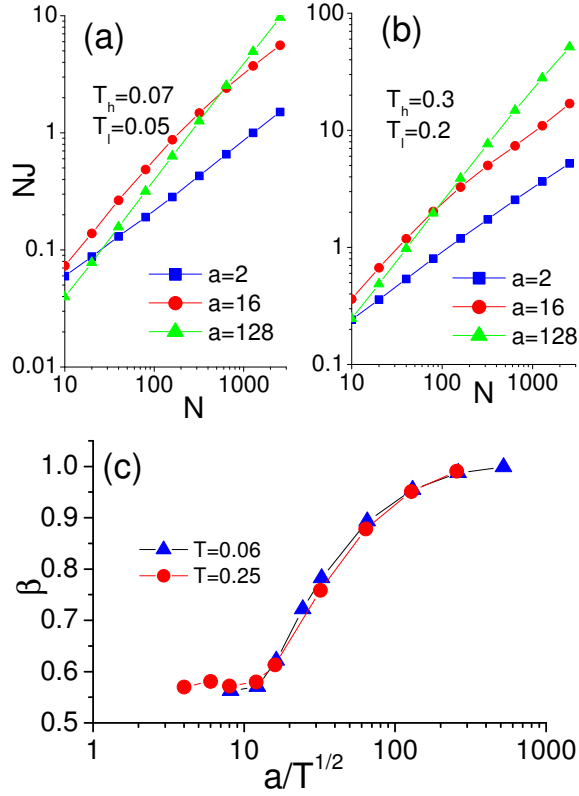}
\end{center}
\caption{$NJ$ versus $N$ for different temperatures. (a)
$T_h=0.07$ and $T_l=0.05$, thus $T\equiv (T_h+T_l)/2=0.06$. (b)
$T_h=0.3$ and $T_l=0.2$, thus $T\equiv (T_h+T_l)/2=0.25$. (c)
$\beta$ versus $a/\sqrt{T}$.} \label{heat_l}
\end{figure}

 The temperature of the {\it i}'th lattice is defined as
$T(i)=(\langle p_{ix}^2\rangle+\langle p_{iy}^2\rangle+\langle
p_{iz}^2\rangle)/3$ and the heat flux along the chain in
stationary state is defined as $J=\langle p_{ix}\frac{\partial
V_{i+1,i}}{\partial x_{i}}+p_{iy}\frac{\partial
V_{i+1,i}}{\partial y_{i}}+ p_{iz}\frac{\partial
V_{i+1,i}}{\partial z_{i}}\rangle$ where
$V_{i+1,i}=\frac{1}{2}(|{\bf r}_{i+1}-{\bf r}_i|-a)^2$. From the
Fourier law $J=-\kappa \nabla T$, the heat
conductivity can be expressed as
   \begin{equation}\label{eq:coefficient}
 \kappa(N)=J(N-1)/(T_h-T_l).
\end{equation}
For fixed $T_h$ and $T_l$, $JN$ can be used to represent the
$\kappa$. For a momentum conservative 1D lattice, it is found that
the heat conduction is anomalous\cite{review,FPU,LW03}, namely,
the heat conductivity depends on the system size as
\begin{equation}\label{eq:alpha}
 \kappa(N)\sim N^{\beta},
\end{equation}
where $\beta$ varies from system to system. In this paper, we
study the dependence of $\beta$ on the lattice constant $a$. For
comparison, we study two concrete cases: a relatively high
temperature regime and a relatively low temperature regime. For
the low temperature case, we take $T_h=0.07$ and $T_l=0.05$, thus
the system temperature defined by $T\equiv (T_h+T_l)/2$ is
$T=0.06$. For high temperature case, we take $T_h=0.3$ and
$T_l=0.2$, thus the system temperature is $T=0.25$ which is about
four times higher than the previous one. The results are shown in
Fig.\ref{heat_l}(a) and (b).
 Our numerical
simulations reveal that $NJ$ increases with $N$ by different
scaling for different $a$ and different temperatures $T_{h/l}$.

%Figure 1

As we have pointed out early that, in the dimensionless
Hamiltonian, the system for different $a$ and different heat bath
temperatures can be equivalent, therefore, in Fig \ref{heat_l}(c)
we draw $\beta$ versus the quantity $a/\sqrt{T}$.

It is quite obvious that two sets of data from Fig.\ref{heat_l}(a)
and (b) almost overlap with each other in Fig.\ref{heat_l}(c). We
have also checked other pairs of $T_h$ and $T_l$ ( such as
$T_h=0.1$ and $T_l=0.005$) and found that they also overlap
with each other. From Fig.\ref{heat_l}(c) it is easy to see that
$\beta$ goes to unity for $a/\sqrt{T}>300$, indicating the large
lattice constant or very low temperature makes the 3D chain
equivalent to a 1D harmonic chain. Moreover, $\beta$ drops with
the decrease of $a$ or increase of temperature, and reaches a
platform for $a/\sqrt{T}\leq 15$, confirming the theoretic
analysis of the previous section.

\begin{figure}[tb]
\begin{center}
\includegraphics{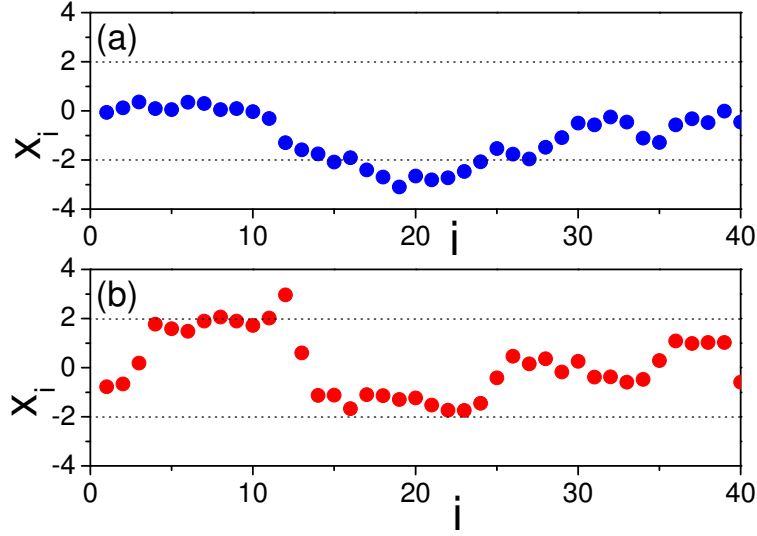}
\end{center}
\caption{Snapshots of $x_i$, the deviation from the equilibrium
position, for $a=2$. (a) $T_h=0.07$ and $T_l=0.05$;
(b) $T_h=0.3$ and $T_l=0.2$. } \label{position}
\end{figure}

We have checked the behaviors of the particles and found that the
particles oscillate around their equilibrium positions and
can be considered as local when $a$ is relatively large and their
oscillation can not be considered as local when $a$ is in the
range of platform. For the relatively small $a$, it is possible
for the particle to move to the area with $|x_i|>a$. Figure
\ref{position} shows the snapshots of the deviation of particle's
positions from their equilibriums. It can be seen that the
positions at $i=17-23$ satisfy $|x_i|>a(=2)$, and the similar
situation occurs in Fig. \ref{position}(b) for $i=12$. This kind
of non-local behavior causes the platform. High
environment temperature causes large deviation  of $|x_i|$ from $a$,
 hence results in a larger platform as  is shown in  Fig. \ref{heat_l}(c).

%Figure 2

\section{Anomalous energy diffusion}

\begin{figure}[tb]
\begin{center}
\includegraphics{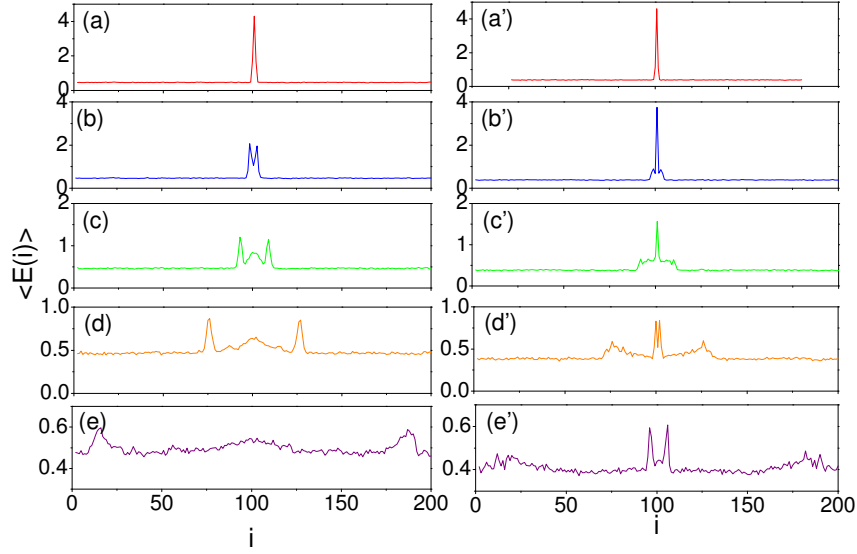}
\end{center}
\caption{Snapshots of $\langle E(i)\rangle$ at
different times for $N=201$, $T=0.2$, and the initial pulse
$p1_{101x}(0)=5p_{101x}(0)$, $p1_{101y}(0)=5p_{101y}(0)$,
$p1_{101z}(0)=5p_{101z}(0)$ is added at the particle $i=101$ at
$t=0$ where the left panels represent the situation of $a=2$, the
right panels represent the situation of $a=128$, and (a) and (a')
denote the snapshot at time $t=1$, (b) and (b') at $t=3$, (c) and
(c') at $t=10$, (d) and (d') at $t=30$, and (e) and (e') at
$t=100$.} \label{snap}
\end{figure}

In order to confirm further the influence of lattice constant on
the heat conduction in the 3D harmonic chain, we study the
diffusion for different $a$. First, we let the two heat baths have
the same temperature and let the chain run enough time to achieve
the equilibrium. Then we give a pulse of energy, which is larger
than the energy of the equilibrium state, at the middle particle
of the chain and investigate how the pulse of energy spreads to
the other part of the chain. Figure \ref{snap} shows the result
for $T_l=T_h=0.2$, $N=201$, and the results are obtained by
averaging over $10^3$ realizations where the initial pulse of five
times the average momentum is added at the particle $i=101$, i.e.,
at $t=0$ we change the velocity of the middle particle to $5$
times of its previous value. Figure \ref{snap}(a) and (a') denote
the snapshot at time $t=1$, (b) and (b') at $t=3$, (c) and (c') at
$t=10$, (d) and (d') at $t=30$, and (e) and (e') at $t=100$, and
the left panels represent the situation of $a=2$ and the right
panels represent the situation of $a=128$.

%Figure 3

Obviously, the left panels of Figure \ref{snap} are different from
the right panels, reflecting the influence of the lattice
constant. Ref. \cite{LW03} gives an approach to measure the energy
diffusion by the formula, $
 \sigma^2(t)=\frac{\int(E(x,t)-E_0)(x-x_0)^2dx}{\int(E(x,t)-E_0)dx},
$ where $E(x,t)$ is the energy distribution at time $t$, $x_0$ is
the position of initial energy pulse at $t=0$, $E_0$ is the energy
of the equilibrium state with temperature $T$. However, in the 3-D
case, this approach does not work very well because the energy
$E(x,t)$ fluctuates around $E_0$ and makes the diffusion
contributions in the integral of this equation cancel each other.

\begin{figure}[tb]
\begin{center}
\includegraphics{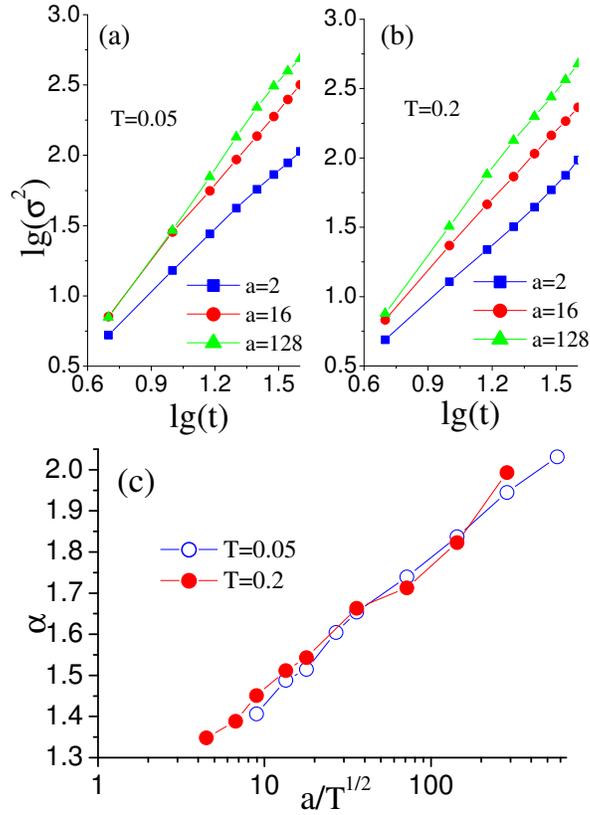}
\end{center}
 \caption{
Three typical situations of $\sigma$ versus $t$ for $T=0.05$ (a)
and $T=0.2$ (b). We have taken an ensemble average over $10^3$
realizations.
 (c) $\alpha$ versus $a/\sqrt{T}$ for $T=0.05$
and $T=0.2$.} \label{diffu_005}
\end{figure}

A correct way is to make the diffusion contributions always
positive, i.e., make the numerator in this equation positive at
each position $x$. Ref.\cite{Cipriani2005} points out recently that, if the pulse is
very small and is initially localized at $i=i_0$, the energy
diffusion can be given by
\begin{equation}\label{eq:diffusion1}
 \sigma^2(t)=\sum_i(i-i_0)^2\delta_{(2)}(i,t),
\end{equation}
where $\delta_{(2)}(i,t)=(\delta v_i(t))^2$ in the 1D case. Here
we extend this approach to the 3D case and take
\begin{eqnarray}\label{eq:delta}
\delta_{(2)}(i,t)&=&(\delta p_{ix}(t))^2+(\delta
p_{iy}(t))^2+(\delta p_{iz}(t))^2 \nonumber \\
& &+(\delta
x_i(t))^2+(\delta y_i(t))^2+(\delta z_i(t))^2.
\end{eqnarray}
That is, $\sigma^2(t)$ measures the total energy variation caused
by the diffusion of the pulse. In numerical simulations, after the
system arrives at the equilibrium we record the time as $t=0$ and
copy all the variables ${\bf x}_i$ and ${\bf p}_i$ to a new set of
variables ${\bf x1}_i$ and ${\bf p1}_i$ except the middle one
${\bf p1}_{101}=1.5{\bf p}_{101}$ where 1.5 chosen so as to make
the initial pulse small enough. Hence, we have two set of systems
(${\bf x}_i, {\bf p}_i$) and (${\bf x1}_i, {\bf p1}_i$) with ${\bf
x1}_i={\bf x}_i$, ${\bf p1}_i={\bf p}_i$ for $i\not=i_0$ and ${\bf
x1}_{i_0}={\bf x}_{i_0}$, ${\bf p1}_{i_0}=1.5{\bf p}_{i_0}$ at
$t=0$. As time goes on, the corresponding variables will gradually
become different because of the diffusion of the pulse. Then, we
can obtain a nonzero $\delta p_{ix}(t)=p1_{ix}(t)-p_{ix}(t)$,
$\delta x_i(t)=x1_i(t)-x_i(t)$, and so on. In this way, we can
find the relation $\sigma^2(t)\sim t^{\alpha}$ through Eqs.
(\ref{eq:delta}) and (\ref{eq:diffusion1}). Figure
\ref{diffu_005}(a) shows the result for three typical values of
$a$ at $T=0.05$, which are obtained by averaging over $10^3$
realizations. The line with ``squares" denotes the case of $a=2$,
the line with ``circles" the case of $a=16$, and the line with
``triangles" the case of $a=128$. We have the similar situations
for other $a$. The slopes of the lines in Fig.\ref{diffu_005}(a)
give the exponent $\alpha$.

%Figure 4

Fig. \ref{diffu_005}(b) show the case of $T=0.2$.

For the same reason as in Fig \ref{heat_l}, we also draw here
$\alpha$ versus $a/\sqrt{T}$ in \ref{diffu_005}(c). The two sets
of date overlap quite well. More interestingly, the Fig.
\ref{diffu_005}(c) clearly indicates that
\begin{equation}
\alpha=A+ B \log_{10} \left(a/\sqrt{T}\right),
 \label{alpha-param}
\end{equation}
where $A\approx 1.138$ and $B\approx 0.325$ are two constants. Why
the exponent of superdiffusion is related to the lattice constant
$a$ and
temperature $T$ in such a relation deserves further investigation.\\

\section{Conclusions and discussions}

In summary, we have studied heat conduction in a quasi 1-D system
with 3D vibration. For finite length (around $1,000$ units or, for
a typical crystal atom chain, $100$ nanometers), the heat
conduction is anomalous and energy diffusion is superdiffusive.
Moreover, we have demonstrated that both the anomalous heat
conduction exponent $\beta$ and the anomalous diffusion exponent
$\alpha$ depends on the dimensionless quantity, $a/\sqrt{T}$. In
particular, we found that $\alpha=A+ B\ln(a/\sqrt{T})$ for reasons to be
further investigated.

We should emphasize that the results obtained in this paper are
for a finite size system, which is more appropriate as the
nanoscale systems are always of finite size. Therefore, the
obtained results may shed lights in understanding heat conduction
in nanoscale structures, especially in networked structure
\cite{LiuLi07}. As the quasi 1D systems such as the nanowires and
nanotubes can be fabricated easily these days, it is therefore of
great interest to do the measurement on the size dependent thermal
conductivity of these systems and to compare with the results
presented.

\section*{Acknowlegments}

This work was partly supported by the National Science Foundation
of China under Grant No. 10775052 and No. 10635040 (ZL) and a
Faculty Research Grant of National University of Singapore.

\end{document}